\documentclass[conference,10pt]{IEEEtran}
\usepackage{color,algorithm,algorithmic,amsbsy,amsmath,amssymb,epsfig,bbm,mathrsfs,fancyhdr,fancyvrb,url}
\usepackage{graphicx}
\usepackage{ulem}
\usepackage{subcaption}
\usepackage{afterpage}  %use this before the definition of the figure --> \afterpage{\clearpage}

\usepackage{xcolor,cite,etoolbox}
\makeatletter
\pretocmd\@bibitem{\color{black}\csname keycolor#1\endcsname}{}{\fail}
\newcommand\citecolor[1]{\@namedef{keycolor#1}{\color{blue}}}
\makeatother
\newcommand{\beq}{\begin{equation}}
\newcommand{\eeq}{\end{equation}}
\newcommand{\beqn}{\begin{eqnarray}}
\newcommand{\eeqn}{\end{eqnarray}}

\newtheorem{lemma}{\textbf{\text{Lemma}}}

\newenvironment{proof}[1][Proof:]{\begin{trivlist}
\item[\hskip \labelsep {\bfseries #1}]}{\end{trivlist}}

\newcommand{\qed}{\nobreak \ifvmode \relax \else
      \ifdim\lastskip<1.5em \hskip-\lastskip
      \hskip1.5em plus0em minus0.5em \fi \nobreak
      \vrule height0.75em width0.5em depth0.25em\fi}
\newcommand\blfootnote[1]{%
  \begingroup
  \renewcommand\thefootnote{}\footnote{#1}%
  \addtocounter{footnote}{-1}%
  \endgroup
}

\begin{document}
\title{Meta Distribution of $\mathrm{SIR}$ in Dual-Hop Internet-of-Things (IoT) Networks}

\author{Hazem Ibrahim, Hina Tabassum, and Uyen T. Nguyen
        % <-this % stops a space
\IEEEcompsocitemizethanks{\IEEEcompsocthanksitem The authors are with the Department of Electrical Engineering and Computer Science, York University, Toronto, Ontario, M3J 1P3 Canada (e-mail:{hibrahim,hina,utn}@cse.yorku.ca).}% <-this % stops an unwanted space
}

%\markboth{Journal of \LaTeX\ Class Files,~Vol.~13, No.~9, September~2014}%
%{Shell\MakeLowercase{\textit{et al.}}: Bare Demo of IEEEtran.cls for Computer Society Journals}
\maketitle
%\IEEEtitleabstractindextext{
\begin{abstract}
This paper characterizes the meta distribution of the downlink signal-to-interference ratio ($\mathrm{SIR}$) attained at a typical Internet-of-Things (IoT) device in a dual-hop IoT network. The IoT device associates with either a serving macro base station (MBS) for direct transmissions or associates with a decode and forward (DF) relay for dual-hop transmissions, depending on the biased received signal power criterion.  In contrast to the conventional success probability, the meta distribution is the distribution of the conditional success probability (CSP), which is  conditioned on the locations of the wireless transmitters. The meta distribution is a fine-grained performance metric that captures important network performance metrics such as the coverage probability and the mean local delay as its special cases. Specifically, we derive the moments of the CSP in order to calculate analytic expressions for the meta distribution. Further, we derive mathematical expressions for special cases such as the mean local delay, variance of the CSP, and success probability of a typical IoT device and typical relay  with different offloading biases. We take in consideration in our analysis the association probabilities of IoT devices. Finally, we investigate the impact of increasing the relay density on the mean local delay using numerical results.

\end{abstract}

%}
\IEEEdisplaynontitleabstractindextext
% \fi
\IEEEpeerreviewmaketitle
\section{Introduction}\label{sec:introduction_report}
{ Internet of Things (IoT) networks are emerging as a key enabling technology for future smart and connected cities,  where a large number of wireless sensors, devices, and actuators will be connected over different wireless technologies\cite{Ericsson}. Generally, devices in an IoT network are battery powered with restricted capacity and are not easily accessible (i.e., due to their deployment in industrial facilities, basements, underground parking lots, and tunnels). As such, highly reliable communication with lower latency will be crucial for the successful deployments of IoT networks \cite{bennis2018ultra}.}
\blfootnote{The authors are with the Department of Electrical Engineering and Computer Science, York University, Toronto, Ontario, M3J 1P3 Canada (e-mail:{hibrahim,hina,utn}@cse.yorku.ca).}

{Existing works on the performance modeling and analysis of IoT networks have focused on characterizing the traditional coverage analysis of M2M communications. Malak et al. \cite{malak2016modeling} studied a generic multihop  uplink communication scheme for M2M  using tools from stochastic geometry. They develop models to characterize the coverage and rate of M2M devices under different transmission schemes. Malak et al. \cite{malak2016optimizing} extended the work in \cite{malak2016modeling} to minimize the energy consumed through a joint optimization of the number of multihop stages and the fraction of aggregators. They propose an energy-efficient data aggregation model for a hierarchical M2M network to optimize the network energy density. Haenggi and Puccinelli \cite{haenggi2005routing} investigate the effect of increasing the numbers of hops on the energy consumption introduced by relays. They concluded that long-hop routing is a competitive strategy but the number of hops cannot be increased arbitrarily due to additional energy consumed by relays.}
\begin{figure}[h]
    \begin{center}
    \scalebox{0.4}[0.4]{\includegraphics{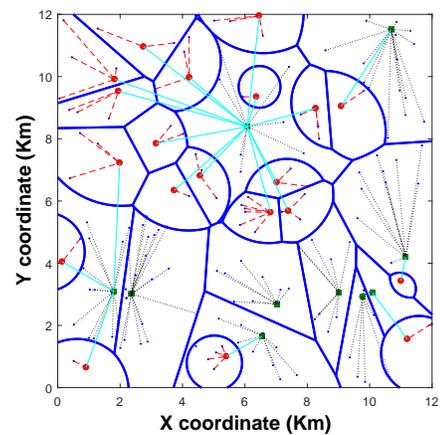}}
    \end{center}
    \caption{\footnotesize{An example of an downlink dual-hop IoT network. MBSs and relays are represented by green squares and red circles, respectively; Devices are represented by blue dots; blue solid lines show cell boundaries; black dotted lines represent direct connections between MBSs and devices; cyan solid lines represent connections between MBSs and relays (first hop of indirect connections); red dashed lines represent connections between relays and devices (second hop of indirect connections)}.
    }
    \vspace{-.2cm}
     \label{stage_2}
    \end{figure}

To date, the performance of dual-hop IoT networks has been characterized in terms of coverage (or success) probability, i.e., the probability that of signal-to-interference-plus noise (SINR) of a channel is greater than the desired SINR threshold. This success probability is the average of the success probabilities over all network realizations. The meta distribution is a more fine-grained metric than the success probability. It provides important metrics such as the standard success probability and mean local delay as its special cases \cite{zhong2016towards}. More formally, the meta distribution can be defined as the distribution of the conditional success probability (CSP), which is conditioned over the locations of the transmitters. That is, the success probability of individual fading links \cite{haenggi2016meta} can be translated to the variability of the communication link reliability.

%Another key performance metric in wireless networks is delay. One of the essential components of the delay is the local delay. The main component of the transmission delay is the retransmission delay which is closely related to the number of retransmissions of a packet which is called local delay \cite{zhong2016towards}.
Fig. \ref{stage_2} depicts the considered dual-hop IoT network. The relays receive transmitted data from MBSs and forward it to IoT devices. These IoT devices are located in isolated areas that do not get coverage from MBSs.  The dual-hop IoT network is targeted to enhance the coverage, energy consumption, transmission delay, and reliability of  IoT devices by receiving and sending data from and to (relatively) closer relays instead of farther MBSs  \cite{kim2018delay,dang2017outage,ibrahim2018data,Hazem2015}.

In this paper, we characterize the meta distribution of the SIR received by a typical IoT device in a dual-hop IoT cellular network  which is composed of two tiers. Tier 1 consisting of macro base stations (MBSs) and  tier 2 made up of low power relays. Specifically, we develop a framework to derive the meta distribution of the SIR of an IoT device that is flexible to associate with a relay (for dual-hop transmissions) or MBS (for direct transmissions) depending on the biased received signal power criterion. We derive closed-form expressions for the moments of the CSP and important network performance metrics such as the mean local delay and the coverage probability. Our numerical results provide useful insights for the reliability and mean local delay of dual-hop IoT networks.

\section{System Model and Assumptions}\label{model}
We consider a downlink cellular wireless network which composed of a two-tier cellular network: tier 1 consisting of MBSs and tier 2 made up of decode and forward (DF) relays. We assume that all relays can communicate with MBSs over wireless connections and they are not connected to the core network. As in \cite{ibrahim2016mobility,Sarabjot2013partition,jo2012heterogeneous,singh2013offloading}, we model the locations of the MBSs and relays as a two-dimensional homogeneous Poisson point process (PPP)  $\mathbf{\Phi}_k=\{\mathbf{y}_{k,1}, \mathbf{y}_{k,2},\cdots\}$ of density $\lambda_{k}$,  where $\mathbf{y}_{k,i}$ is the location of  $i^{\mathrm{th}}$ MBS (when $k=1$) or the $i^{\mathrm{th}}$ relay (when $k=2$).  Let $\mathcal{D}$ denote the set of devices. The locations of devices in the network are modeled as an independent homogeneous PPP  $\mathbf{\Phi}_\mathcal{D}=\{\mathbf{x}_{1}, \mathbf{x}_{2},\cdots\}$ with density $\lambda_{\mathcal{D}}$, where $\mathbf{x}_{i}$ is the location of the $i^{\mathrm{th}}$ device.
We assume that the intensity $\lambda_\mathcal{D}$ of IoT devices is much higher than the intensity of MBSs and relays. MBSs and relays are considered to operate on orthogonal spectrum; thus, there is no inter-tier interference. Let $W$ denotes the total available spectrum in the network, given by $W=W_1+W_2$, where the spectrum allocated to MBSs is given by $W_{1} = \eta W$, and the spectrum allocated to relays is given by $W_{2} =(1-\eta)W$, respectively.

{Each IoT device associates with either a MBS or a relay depending on the maximum biased received  power which is measured at devices. On the other hand, each relay associates with the MBS that offers the maximum received power which is measured at relays. The association criterion for an IoT device can then be described mathematically as follows \cite{jo2012heterogeneous}:}
      \begin{equation}\label{biasing}
    P_{k}B_{k}(\min_{i}\|\mathbf{y}_{k,i}-\mathbf{x}\|)^{-\alpha_{k}}\ge P_{j}B_{j}(\min_{i^{'}}\|\mathbf{y}_{j,i^{'}}-\mathbf{x}\|)^{{-\alpha_{k}}}, \forall j
    \end{equation}
    \noindent
    where $\|.\|$ denotes the Euclidean distance, $B_{1}$ and $B_{2}$ are the association bias for MBSs and relays, respectively. The association bias $B_{2}$ artificially encourages/discourages IoT devices to associate with the relays \cite{jo2012heterogeneous}. Note that we do not consider association bias between relays and MBSs. Therefore,  a relay associates with a MBS based with the maximum received power in the downlink, which is measured at relays, i.e.,  $B_{1}=1$. A device or relay associates with a serving node (given by Eq. (\ref{biasing})), which is termed \emph{\textbf{tagged node}} (tagged relay or tagged MBS).

A general power law path loss model is considered in which the signal power decays at rate $d^{-\alpha_{k}}$ with the distance $d$, where $\alpha_{k}$ is the path loss exponent for tier 1 when $k=1$ and for tier 2 when $k=2$ and $\alpha_{k}>2$. Let $h(\mathbf{x}, \mathbf{y})$ denotes the channel gain (fast fading) between two generic locations $\mathbf{x}$, $\mathbf{y}$ $\in  \mathbb{R}$. The channel gain $h(\mathbf{x}, \mathbf{y})\sim \exp(1)$ is exponentially distributed with unit mean (Rayleigh fading with power normalization). All wireless nodes in the $k^{th}$ tier transmit with the same transmit power $P_{k}$ in the downlink. Let an IoT device be located at $\mathbf{x}$ and associated with a tagged node (relay or MBS) at location $\mathbf{y}$. The received power at the device can then be given as $P_{k}\|\mathbf{y}-\mathbf{x}\|^{-\alpha_{k}}h(\mathbf{x}, \mathbf{y})$, based on the associated tier.

 For the sake of clarity, we define $\hat{P}_{jk}\overset{\Delta}{=}\frac{P_{j}}{P_{k}}$, $\hat{B}_{jk}\overset{\Delta}{=}\frac{B_{j}}{B_{k}}$, $\hat{\lambda}_{jk}\overset{\Delta}{=}\frac{\lambda_{j}}{\lambda_{k}}$. As derived in \cite{jo2012heterogeneous}, the conditional association probability for the typical device connecting to the $k^{th}$ tier (conditional over the desired link distance $d_{\mathcal{D},k}$) is as follows:
    \begin{equation}\label{condition-access}
      \mathbb{P}(n=k|d_{\mathcal{D},k})=\prod_{j\neq k}e^{-\pi\lambda_{j}(\hat{P}_{jk}\hat{B}_{jk})^{2/\alpha_{j}}r^{2}},
    \end{equation}
  where $n$ denotes an index of tier which  the typical device associating with.

%A two-hop IoT network can be constructed from any configuration of the three PPPs $\mathbf{\Phi}_1$, $\mathbf{\Phi}_2$, and $\mathbf{\Phi}_\mathcal{D}$, where each device of the PPP $\mathbf{\Phi}_\mathcal{D}$ associates with either a relay from PPP $\mathbf{\Phi}_2$ or with a MBS from PPP $\mathbf{\Phi}_1$ based on the maximum biased downlink received signal power from either MBSs or relays. {Let $L_{1,\mathcal{D}}$ denotes the direct link when  a device associates  with MBSs directly for downlink transmission. On the other hand, indirect connection occurs when a given IoT device connects to a  relay. Let $L_{1,2}$ denotes the \emph{first hop} of the indirect connection (from MBSs to relays) and   $L_{2,\mathcal{D}}$ denotes the \emph{second hop} of the indirect connection (from relays to devices).  The downlink transmission goes through two consecutive links $L_{1,2}$ and $L_{2,\mathcal{D}}$; Therefore, the link with the minimum data rate becomes the bottleneck and determines the end-to-end throughput of the indirect downlink connection. Fig. \ref{stage_2} demonstrates the considered dual-hop IoT network.}

\section{Definitions, SIR Models and Methodology of Analysis}

\subsection{The Meta Distribution: Definitions and Calculations}
The {meta distribution} $\bar{F}_{P_{s}}(x)$ is the complementary cumulative distribution function (CCDF) of the conditional success probability (CSP) $P_{s}(\theta)$ and it is given by \cite{haenggi2016meta}:
          \begin{equation}\label{meta-1}
            \bar{F}_{P_{s}}(x)\overset{\Delta}{=}\mathbb{P}(P_{s}(\theta)>x), \quad\quad x\in [0,1],
          \end{equation}
         where, conditioned on the locations of the transmitters
and that the desired transmitter is active, CSP $P_{s}(\theta)$ of a typical device \cite{haenggi2016meta} can be defined as follows:
          \begin{equation}\label{meta}
            P_{s}(\theta)\overset{\Delta}{=}\mathbb{P}(\mathrm{SIR}>\theta|\mathbf{\Phi},\text{tx})
          \end{equation}
          where $\theta$ is the desired $\mathrm{SIR}$. Physically, the meta distribution provides the fraction of the active links whose CSPs are greater than $x$, since all point processes in the model are ergodic.

Let $M_{b}$ denotes the $b^{th}$ moment of $P_{s}(\theta)$, i.e., $M_{b}\overset{\Delta}{=}\mathbb{E}^{0}(P_{s}(\theta)^{b})$, $b\in\mathbb{C}$.
The \textbf{exact meta distribution} can be calculated by using the Gil-Pelaez theorem \cite{gil1951note} after deriving the imaginary moments of $P_{s}(\theta)$, i.e., $M_{jt}$, $j\overset{\Delta}{=}\sqrt{-1}$, $t\in\mathbb{R}^{+}$. By applying the Gil-Pelaez theorem \cite{gil1951note} to the imaginary moments $M_{jt}$, where $t\in\mathbb{R}$, we get the exact meta distribution of the total network as follows:
  \begin{equation}\label{Gil-pelaez-2hops}
    \bar{F}_{P_{s,\text{total}}}(x)=\frac{1}{2}+\frac{1}{\pi}\int_{0}^{\infty}\frac{\Im\left(e^{-jt\log{x}}M_{jt}\right)}{t}\text{d}t,
  \end{equation}
  where $\Im(w)$ is imaginary part of $w\in \mathbb{C}$.

      Using moment matching techniques, the \textbf{meta distribution of the CSP} can be approximated using the Beta distribution as follows:
     \begin{equation}\label{beta-match}
       \bar{F}_{P_{s}}(x)\approx 1-I_{x}\left(\frac{\beta M_{1}}{1-M_{1}},\beta\right), \quad\quad x\in [0,1],
     \end{equation}
     where \scriptsize$\beta\overset{\Delta}{=}\frac{(M_{1}-M_{2})(1-M_{1})}{M_{2}-M_{1}^{2}}$\normalsize; $M_{1}$ and $M_{2}$ represent the first and the second moments, respectively;
     and $I_{x}(a,b)$ is the regularized incomplete Beta function
     \begin{equation}\label{regularized-incomplete-beta-function}
       I_{x}(a,b)\overset{\Delta}{=}\frac{\int_{0}^{x}t^{a-1}(1-t)^{b-1}d\text{t}}{B(a,b)},
     \end{equation}
     where $B(a,b)$ denotes the Beta function.
\subsection{SIR Models for Dual Hop IoT Networks}
The considered dual-hop IoT network is comprised of
\begin{itemize}
    \item Direct transmission links from MBSs to IoT devices.
    \item Dual-hop transmission links with a first hop link from MBSs to relays and second hop links from relays to IoT devices.
\end{itemize}

We use subscript ``$1,2$'', ``$2,\mathcal{D}$'', ``$1,\mathcal{D}$'', and ``$\mathcal{D}$'' to denote first hop links, second hop links, direct  links, and devices, respectively. Let $\theta_{2}$ and $\theta_{\mathcal{D}}$ denote the predefined threshold for correct signal reception at a typical relay from a MBS (first hop) and at a typical device from either a relay (second hop) or from a MBS (direct connection), respectively.
\subsubsection{First Hop}
The $\mathrm{SIR}$ of a typical relay associated with a MBS can be modeled as follows:
      \small
       \begin{equation} \label{SIR-backhaul-O}
       \mathrm{SIR}_{1,2} = \frac {P_{1}d_{1,2}^{-\alpha_{1}}h(0, \mathbf{y}_{1,0})}{\mathcal{I}_{1,2}},
       \end{equation}
       \normalsize
       where $\mathcal{I}_{1,2}$ denote the total interference from the set of active MBSs $\mathbf{\Phi}_{1}$ scheduled to transmit on the same resource block, excluding the tagged MBS. Then $\mathcal{I}_{1,2}$ is calculated as follows:
       \begin{equation}
        \mathcal{I}_{1,2}=P_{1}\sum_{i:\mathbf{y}_{1,i}\in\mathbf{\Phi}_1\backslash \{\mathbf{y}_{1,0}\}}\|\mathbf{y}_{1,i}\|^{-\alpha_{1}}h(0,\mathbf{y}_{1,i}).
        \end{equation}

\subsubsection{Second Hop} The $\mathrm{SIR}$ of a typical device associated with a relay can be modeled as follows:
         \begin{equation} \label{SIR-access-O}
       \mathrm{SIR}_{2,\mathcal{D}} = \frac {P_{2}d_{2,\mathcal{D}}^{-\alpha_{2}}h(0, \mathbf{y}_{2,0})}{\mathcal{I}_{2,\mathcal{D}}},
       \end{equation}
       where $\mathcal{I}_{2,\mathcal{D}}$ denote the total interference from the set of active relays $\mathbf{\Phi}_{2}$ scheduled to transmit on the same resource block, excluding the tagged relay. Then $\mathcal{I}_{2,\mathcal{D}}$ is calculated as follows:
       \begin{equation}
        \mathcal{I}_{2,\mathcal{D}}=P_{2}\sum_{i:\mathbf{y}_{2,i}\in\mathbf{\Phi}_2\backslash \{\mathbf{y}_{2,0}\}}\|\mathbf{y}_{2,i}\|^{-\alpha_{2}}h(0,\mathbf{y}_{2,i}).
        \end{equation}
         The PDF of the distance $d_{1,2}$ between the typical relay and its tagged MBS can be given as follows \cite{andrews2011tractable}:
  \begin{equation}\label{one_tier_distance}
  f_{d_{1,2}}=2\pi\lambda_{1}re^{-\lambda_{1}\pi r^{2}}.
\end{equation}

\subsubsection{Direct Connection} The $\mathrm{SIR}$ of a typical device associated directly with a MBS is as follows:
       \begin{equation} \label{SIR-direct-link}
       \mathrm{SIR}_{1,\mathcal{D}} = \frac{P_{1}d_{1,\mathcal{D}}^{-\alpha_{1}}h(0, \mathbf{y}_{1,0})}{\mathcal{I}_{1,\mathcal{D}}},
       \end{equation}
       where $\mathcal{I}_{1,\mathcal{D}}$ denote the total interference from the set of active MBSs $\mathbf{\Phi}_{1}$ scheduled to transmit on the same resource block for a typical device, excluding the tagged MBS. Then $\mathcal{I}_{1,\mathcal{D}}$ is calculated as follows:
       \begin{equation}
        \mathcal{I}_{1,\mathcal{D}}=P_{1}\sum_{i:\mathbf{y}_{1,i}\in\mathbf{\Phi}_1\backslash \{\mathbf{y}_{1,0}\}}\|\mathbf{y}_{1,i}\|^{-\alpha_1}h(0,\mathbf{y}_{1,i}).
        \end{equation}

\subsection{Methodology of Analysis}
Since the success probability of a dual-hop transmission depends on the success probabilities of transmissions on both hops, the meta distribution of the  dual-hop link can be defined as follows:
\begin{equation}\label{meta-2}
            \bar{F}_{P_{s,\text{dual-hop}}}(x)\overset{\Delta}{=}\mathbb{P}^{0}(P_{s,\mathrm{FH}}(\theta_{2})P_{s,2}(\theta_{\mathcal{D}})>x), \quad\quad x\in [0,1],
          \end{equation}
where
\begin{equation}\label{meta_FS_2}
            P_{s,\mathrm{FH}}(\theta_{2})\overset{\Delta}{=}\mathbb{P}(\mathrm{SIR_{1,2}}>\theta_{2}|\mathbf{\Phi_1},\text{tx}),
          \end{equation}
          and
          \begin{equation}\label{meta_second_hop_2}
            P_{s,2}(\theta_{\mathcal{D}})\overset{\Delta}{=}\mathbb{P}(\mathrm{SIR_{2,\mathcal{D}}}>\theta_{\mathcal{D}}|\mathbf{\Phi}_2,\text{tx}).
          \end{equation}
          Note that the CSPs $P_{s,\mathrm{FH}}(\theta_{2})$ and $P_{s,2}(\theta_{\mathcal{D}})$ are independent random variables, conditioned on the locations of the transmitters. Also, as we assume orthogonal spectrum allocation on the two hops, the two transmissions do not interfere with each other. Thus, there is no correlation between the two hops. Subsequently, our methodology of analysis can be given as follows:
          \begin{itemize}
          \item Derive the CSPs $P_{s,1}(\theta_{\mathcal{D}})$ and $P_{s,2}(\theta_{\mathcal{D}})$ as in Eq. (\ref{MBS-OR-SBS}).
          \item Derive the $b^{th}$ moment of the CSP $P_{s,k}(\theta_{\mathcal{D}})$ of the typical device when it is served by the $k^{th}$ tier, i.e., $M_{b,k} \quad \forall k=1,2$, which is defined as \small$\mathbb{E}_{d_{k,\mathcal{D}}}\bigg[\mathbb{P}(n=k|d_{k,\mathcal{D}})P_{s,k}(\theta_{\mathcal{D}})^{b}\bigg]$\normalsize  as shown in Lemma \ref{1}. Note that we take the offloading bias effect into consideration while calculating the $b^{th}$ moment of the CSP $P_{s,k}(\theta_{\mathcal{D}})$.
          \item Derive the CSP $P_{s,\mathrm{FH}}(\theta_{\mathcal{D}})$ of the first hop link in Eq. (\ref{P_s_b}).
          \item Derive the $b^{th}$ moment of $P_{s,\mathrm{FH}}(\theta_{\mathcal{D}})$ in Eq. (\ref{two-hops-b-backhaul}).
          \item Derive the total network moment of the CSP in Eq.(\ref{whole-moments})
          \small
\begin{align}\label{total_meta}
        M_{b,\text{total}}&= \underbrace{M_{b,\text{dual-hop}}}_\text{Dual-hop transmission}+\underbrace{M_{b,1}}_\text{Direct transmission}= M_{b,FH} M_{b,2}+M_{b,1}.
\end{align}
\normalsize
\item Apply the Gil-Pelaez theorem using the total network moment $M_{b,\text{total}}$ to get the exact meta distribution of the total network as shown in Eq. (\ref{Gil-pelaez-2hops}).
    \item Approximate the meta distribution of the CSP of the whole network by using the Beta distribution as in Eq. (\ref{beta-match}).
          \end{itemize}

For the first hop link analysis, we condition on having a typical device at the origin, which becomes the typical device under expectation over the point processes due to the stationary nature of the PPPs \cite{haenggi2012book}. Similarly, for the second hop,  we condition on having a relay at the origin which becomes the typical relay under expectations over the point processes.
\section{Analysis of the Meta Distribution}
      The typical device at the origin associates with the MBS tier (when $k=1$) or the relay tier (when $k=2$), given $d_{k,\mathcal{D}}$. Using the expressions of  $\mathrm{SIR}_{2,\mathcal{D}}$ and $\mathrm{SIR}_{1,\mathcal{D}}$ calculated in Eq. (\ref{SIR-access-O}) and Eq. (\ref{SIR-direct-link}), respectively, we calculate a general expression for \textbf{the CSP of the second hop link} $P_{s,k}(\theta_{\mathcal{D}})$ (when $k=2$) and \textbf{the CSP of the direct connection} (when $k=1$). The CSP for both cases are $P_{s,1}(\theta_{\mathcal{D}})$ and $P_{s,2}(\theta_{\mathcal{D}})$, respectively. We calculate $P_{s,1}(\theta_{\mathcal{D}})$ and $P_{s,2}(\theta_{\mathcal{D}})$ by substituting Eq. \eqref{SIR-direct-link} and Eq. \eqref{SIR-access-O} in \eqref{meta} as follows \cite{haenggi2016meta}:
      \small
    \begingroup
    \allowdisplaybreaks
     \begin{align}\label{MBS-OR-SBS}
        &P_{s,k}(\theta_{\mathcal{D}})\notag\\
        &=\mathbb{P}\left(h(0, \mathbf{y}_{k,0})>\frac{\theta_{\mathcal{D}}d_{k,\mathcal{D}}^{\alpha_{k}}}{P_{k}}\mathcal{I}_{k,\mathcal{D}}|\mathbf{\Phi}_{1},\mathbf{\Phi}_{2},\text{tx}\right),\notag\\
        &\stackrel{(a)}{=}\mathbb{E}\bigg[\exp\left(-\theta_{\mathcal{D}}d_{k,\mathcal{D}}^{\alpha_{k}}\sum_{i:\mathbf{y}_{k,i}\in\mathbf{\Phi}_k\backslash \{\mathbf{y}_{k,0}\}}\|\mathbf{y}_{k,i}\|^{-\alpha_{k}}h(0,\mathbf{y}_{k,i})\right)\bigg],\notag\\
        &=\prod_{\mathbf{y}_{k,i}\in\mathbf{\Phi}_k\backslash \{\mathbf{y}_{k,0}\}}\mathbb{E}\bigg[\exp\left(-\theta_{\mathcal{D}}d_{k,\mathcal{D}}^{\alpha_{k}}\|\mathbf{y}_{k,i}\|^{-\alpha_{k}}h(0,\mathbf{y}_{k,i})\right)\bigg]\notag,\\
        &\stackrel{(b)}{=}\prod_{\mathbf{y}_{k,i}\in\mathbf{\Phi}_k\backslash \{\mathbf{y}_{k,0}\}}\frac{1}{1+\theta_{\mathcal{D}}d_{k,\mathcal{D}}^{\alpha_{k}}\|\mathbf{y}_{k,i}\|^{-\alpha_{k}}},\notag\\
        &=\prod_{\mathbf{y}_{k,i}\in\mathbf{\Phi}_k\backslash \{\mathbf{y}_{k,0}\}}\frac{1}{1+\theta_{\mathcal{D}}\left(\frac{d_{k,\mathcal{D}}}{\|\mathbf{y}_{k,i}\|}\right)^{\alpha_k}}.
      \end{align}
      \endgroup
      \normalsize
      where (a) follows from the fact that the channel gain $h(0, \mathbf{y}_{k,0})\sim \exp(1)$ is independently exponentially distributed with unit mean and (b) is obtained by taking the expectation with respect to $h(0,\mathbf{y}_{k,i})$. Therefore, the $b^{th}$ moment of $P_{s,k}(\theta_{\mathcal{D}})$ of the typical device when it is served by the $k^{th}$ tier is characterized by the following lemma.
    \begin{lemma}[The \textbf{$b^{th}$ moment of $P_{s,k}(\theta_{\mathcal{D}})$ of the typical device when it is served by the $k^{th}$ tier.}] \label{1}
    While taking the offloading bias effect into consideration, the $b^{th}$ moment of the typical device when it is served by the $k^{th}$ tier is as follows:\small
          \begin{align}\label{MBS-or-SBS-b-moment}
          M_{b,\text{k}}&=\frac{1}{\sum\limits_{j\neq k}\hat{\lambda}_{jk}(\hat{P}_{jk}\hat{B}_{jk})^{2/\alpha_{j}}+\text{ }_{2}F_{1}(b,-\frac{2}{\alpha_{k}};1-\frac{2}{\alpha_{k}};-\theta_{\mathcal{D}})}.
          \end{align}
    \normalsize
    \end{lemma}
\vspace{-0.8cm}
\begin{proof}
See Appendix \ref{prof_1}.
\end{proof}

 First hop (FH) link: Similarly, using $\mathrm{SIR}_{1,2}$ defined in Eq. (\ref{SIR-backhaul-O}), we define the \textbf{CSP of the first hop link} by substituting Eq. \eqref{SIR-backhaul-O} into Eq. \eqref{meta} as follows:
    \small
       \begin{align}\label{P_s_b}
        &P_{s,\mathrm{FH}}(\theta_{2})\notag\\
        &=\mathbb{P}\left(h(0, \mathbf{y}_{1,0})>\frac{\theta_{2}d_{1,2}^{\alpha_1}}{P_{1}}\mathcal{I}_{1,2}|\mathbf{\Phi}_{1},\mathbf{\Phi}_{2},\text{tx}\right),\notag\\
        &\stackrel{(a)}{=}\mathbb{E}\bigg[\exp\left(-\theta_{2}d_{1,2}^{\alpha_1}\sum_{i:\mathbf{y}_{1,i}\in\mathbf{\Phi}_1\backslash \{\mathbf{y}_{1,0}\}}\|\mathbf{y}_{1,i}\|^{-\alpha_1}h(0,\mathbf{y}_{1,i})\right)\bigg],\notag\\
        &=\prod_{\mathbf{y}_{1,i}\in\mathbf{\Phi}_1\backslash \{\mathbf{y}_{1,0}\}}\mathbb{E}\bigg[\exp\left(-\theta_{2}d_{1,2}^{\alpha_1}\|\mathbf{y}_{1,i}\|^{-\alpha_1}h(0,\mathbf{y}_{1,i})\right)\bigg]\notag,\\
        &\stackrel{(b)}{=}\prod_{\mathbf{y}_{2,i}\in\mathbf{\Phi}_1\backslash \{\mathbf{y}_{1,0}\}}\frac{1}{1+\theta_{2}d_{1,2}^{\alpha_1}\|\mathbf{y}_{1,i}\|^{-\alpha_1}},\notag\\
        &=\prod_{\mathbf{y}_{1,i}\in\mathbf{\Phi}_1\backslash \{\mathbf{y}_{1,0}\}}\frac{1}{1+\theta_{2}\left(\frac{d_{1,2}}{\|\mathbf{y}_{1,i}\|}\right)^{\alpha_1}}.
      \end{align}
      \normalsize
where (a) follows from the fact that the channel gain $h(0, \mathbf{y}_{1,0})\sim \exp(1)$ is independently exponentially distributed with unit mean and (b) is obtained by taking the expectation with respect to $h(0,\mathbf{y}_{2,i})$. Therefore, the \textbf{$b^{th}$ moment of the first hop}, i.e., between a typical relay and the MBS it associates with, is given by:
\small
\begin{align}\label{two-hops-b-backhaul}
           M_{b,\mathrm{FH}}&=\mathbb{E}\bigg[P_{s,\mathrm{FH}}(\theta_{2})^{b}\bigg],\notag\\
                 &=\mathbb{E}\bigg[\prod_{\mathbf{y}_{1,i}\in\mathbf{\Phi}_1\backslash \{\mathbf{y}_{1,0}\}}\frac{1}{\left(1+\theta_{2}\left(\frac{d_{1,2}}{\|\mathbf{y}_{1,i}\|}\right)^{\alpha_1}\right)^{b}}\bigg],\notag\\
                 &=\left(\frac{1}{1+2\int_{0}^{1}\left(1-\frac{1}{(1+\theta_{2}r^{\alpha_1})^{b}}\right)r^{-3}dr}\right),\notag\\
                 &=\frac{1}{\text{ }_{2}F_{1}(b,-\frac{2}{\alpha_1};1-\frac{2}{\alpha_1};-\theta_{2})},
          \end{align}
          \normalsize
         where (a) follows from Lemma 1 in \cite{ganti2016asymptotics}. This lemma calculates the probability generating functional (PGFL) of the relative distance process (RDP) defined as $\mathcal{R}\overset{\Delta}{=}\{x\in\mathbf{\Phi}\backslash \{\mathbf{x}_0\}:\|x_0\|/\|x\|\}\subset(0,1)$, when $\mathbf{\Phi}$ is a PPP. Therefore, we apply the PGFL of the RDP as follows
         \small
         \begin{equation}\label{pgfl-RDP}
           G_{\mathcal{R}}[f]\overset{\Delta}{=}\mathbb{E}\prod_{x\in\mathcal{R}}f(x)=\frac{1}{1+2\int_{0}^{1}(1-f(x))x^{-3}dx}.         \end{equation}
           \normalsize

\subsection{The $b^{th}$ Moment of Dual-Hop Links with Offloading Biases}
We use the $b^{th}$ moment of the first hop and the $b^{th}$ moment of the second hop defined in Eq. \eqref{two-hops-b-backhaul} and Eq. \eqref{MBS-or-SBS-b-moment}, respectively, to calculate the $b^{th}$ moment of the dual-hop link with offloading biases, by taking the expectation over the point processes as follows:
          \small
          \begin{align}\label{two-hops-b-moment}
            &M_{b,\text{dual-hop}}\notag\\
            &=\mathbb{E}\bigg[P_{s,\mathrm{FH}}(\theta_{2})^{b}\times \prod_{j\neq k}e^{-\pi\lambda_{j}(\hat{P}_{jk}\hat{B}_{jk})^{2/\alpha_{j}}r^{2}} P_{s,2}(\theta_{\mathcal{D}})^{b}\bigg],\notag\\
                 &\stackrel{(a)}{=}\underbrace{\mathbb{E}\bigg[P_{s,\mathrm{FH}}(\theta_{2})^{b}\bigg]}_\text{$M_{b,\mathrm{FH}}$(first hop )}\underbrace{\mathbb{E}\bigg[\prod_{j\neq k}e^{-\pi\lambda_{j}(\hat{P}_{jk}\hat{B}_{jk})^{2/\alpha_{j}}r^{2}}P_{s,2}(\theta_{\mathcal{D}})^{b}\bigg]}_\text{$M_{b,2}$(second hop)},\notag\\
                 &\stackrel{(b)}{=}\frac{1}{\text{ }_{2}F_{1}(b,-\frac{2}{\alpha_1};1-\frac{2}{\alpha_1};-\theta_{2})}\times\notag\\
                 &\frac{1}{\hat{\lambda}_{12}(\hat{P}_{12}\hat{B}_{12})^{2/\alpha_{1}}+\text{ }_{2}F_{1}(b,-\frac{2}{\alpha_{2}};1-\frac{2}{\alpha_{2}};-\theta_{\mathcal{D}})}
          \end{align}
          \normalsize
         where (a) follows from the independency between the locations of the MBSs and relays which are modeled as  two-dimensional independent homogenous PPPs on $\mathbb{R}^2$, $\mathbf{\Phi}_1$ and $\mathbf{\Phi}_2$, respectively. In step (b) we substitute the definition of $M_{b,\text{FH}}$ from Eq. \eqref{two-hops-b-backhaul} and the definition of $M_{b,\text{2}}$ from Eq. \eqref{MBS-or-SBS-b-moment} when $k=2$ into the total meta distribution for the entire dual-hop IoT network represented by Eq. (\ref{total_meta}). Therefore, using Eq. \eqref{MBS-or-SBS-b-moment} (when $k=1$) and Eq. \eqref{two-hops-b-moment}, we get the total network $b^{th}$ moment as shown at the top of the next page in Eq. (\ref{whole-moments}).
         \begin{figure*}[ht]
\footnotesize
  \begin{align}\label{whole-moments}
        M_{b,\text{total}}
        &=\frac{1}{\text{ }_{2}F_{1}(b,-\frac{2}{\alpha_1};1-\frac{2}{\alpha_1};-\theta_{2})}\frac{1}{\hat{\lambda}_{12}(\hat{P}_{12}\hat{B}_{12})^{2/\alpha_{1}}+\text{ }_{2}F_{1}(b,-\frac{2}{\alpha_{2}};1-\frac{2}{\alpha_{2}};-\theta_{\mathcal{D}})}+\frac{1}{\hat{\lambda}_{21}(\hat{P}_{21}\hat{B}_{21})^{2/\alpha_{2}}+\text{ }_{2}F_{1}(b,-\frac{2}{\alpha_{1}};1-\frac{2}{\alpha_{1}};-\theta_{\mathcal{D}})}.
      \end{align}
      \hrule
\normalsize
\end{figure*}
  \begin{figure*}[ht]
          \footnotesize
          \begin{flushleft}
         \begin{align}\label{first-momemnt-beta}
           M_{1,\text{total}}&=\frac{1}{\text{ }_{2}F_{1}(1,-\frac{2}{\alpha_1};1-\frac{2}{\alpha_1};-\theta_{2})}\frac{1}{\hat{\lambda}_{12}(\hat{P}_{12}\hat{B}_{12})^{2/\alpha_{1}}+\text{ }_{2}F_{1}(1,-\frac{2}{\alpha_{2}};1-\frac{2}{\alpha_{2}};-\theta_{\mathcal{D}})}
        +\frac{1}{\hat{\lambda}_{21}(\hat{P}_{21}\hat{B}_{21})^{2/\alpha_{2}}+\text{ }_{2}F_{1}(1,-\frac{2}{\alpha_{1}};1-\frac{2}{\alpha_{1}};-\theta_{\mathcal{D}})},
         \end{align}
         \begin{align}\label{second-momemnt-beta}
           M_{2,\text{total}}&=\frac{1}{\text{ }_{2}F_{1}(2,-\frac{2}{\alpha_1};1-\frac{2}{\alpha_1};-\theta_{2})}\frac{1}{\hat{\lambda}_{12}(\hat{P}_{12}\hat{B}_{12})^{2/\alpha_{1}}+\text{ }_{2}F_{1}(2,-\frac{2}{\alpha_{2}};1-\frac{2}{\alpha_{2}};-\theta_{\mathcal{D}})}+
        \frac{1}{\hat{\lambda}_{21}(\hat{P}_{21}\hat{B}_{21})^{2/\alpha_{2}}+\text{ }_{2}F_{1}(2,-\frac{2}{\alpha_{1}};1-\frac{2}{\alpha_{1}};-\theta_{\mathcal{D}})}.
         \end{align}
         \end{flushleft}
         \normalsize
         \hrule
         \end{figure*}

\subsection{The Exact Meta Distribution}
   By applying the Gil-Pelaez theorem \cite{gil1951note} to the imaginary moments $M_{jt}$, where $t\in\mathbb{R}$, we get the exact meta distribution of the total network as follows:
  \small
  \begin{equation}\label{Gil-pelaez-2hops}
    \bar{F}_{P_{s,\text{total}}}(x)=\frac{1}{2}+\frac{1}{\pi}\int_{0}^{\infty}\frac{\Im\left(e^{-jt\log{x}}M_{jt,\text{total}}\right)}{t}\text{d}t,
  \end{equation}
  \normalsize
  where $\Im(w)$ is the imaginary part of $w\in \mathbb{C}$.

   \begin{figure}[!th]
    \begin{center}
    \scalebox{0.5}[0.5]{\includegraphics{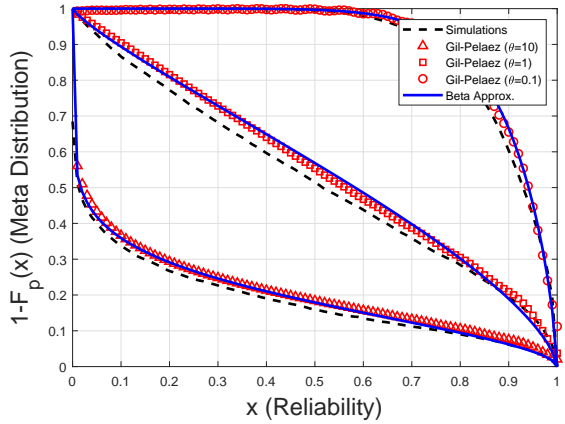}}
    %\includegraphics[width=\textwidth,keepaspectratio]{./figures/untitled.jpg}
    %\vspace{-.5cm}
    \end{center}
    \caption{The meta distribution as a function of reliability x for $\theta=\theta_{\mathcal{D}}=\theta_{2}=$10, 1, and 0.1 when $B_{1}=B_{2}=1$ and $\alpha_{1}=\alpha_{2}=4$.}
     \label{meta_selfbackhaul}
    \end{figure}

     \begin{figure}[!th]
    \begin{center}
    \scalebox{0.4}[0.4]{\includegraphics{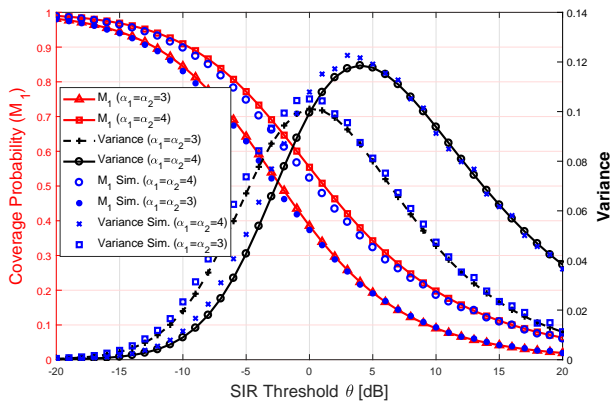}}
    %\includegraphics[width=\textwidth,keepaspectratio]{./figures/untitled.jpg}
    %\vspace{-.5cm}
    \end{center}
    \caption{Coverage probability $M_{1,\text{total}}$ and variance $M_{2,\text{total}}-M_{1,\text{total}}^{2}$ as a function of $\theta$ when $B_{1}=B_{2}=1$ and $\alpha_{1}=\alpha_{2}=3$ and 4.}
     \label{coverage_variance}
    \end{figure}

\subsection{Beta Approximation}
The meta distribution expression derived using the Gil-Pelaez in \eqref{Gil-pelaez-2hops} is exact and it is difficult to evaluate numerically since it usually involves many calculations of imaginary moments, which prohibits direct insights into the meta distribution and its applications in mapping to other performance metrics like the ergodic data rate \cite{deng2017fine}. Therefore, we follow \cite{haenggi2016meta,salehi2018meta,wang2017meta,wang2018SIR} to approximate the meta distribution by a Beta distribution by matching the first and second moments, which are easily obtained from the general result in Eq. \eqref{whole-moments}. The results are shown in Eq. (\ref{first-momemnt-beta}) and (\ref{second-momemnt-beta}) at the top of this page.

         \begin{figure*}[ht]
\footnotesize
      \begin{align}\label{local_delay_total}
        &M_{-1,\text{total}}\notag\\&=\frac{1}{\text{ }_{2}F_{1}(-1,-\frac{2}{\alpha_1};1-\frac{2}{\alpha_1};-\theta_{bh})}\frac{1}{\hat{\lambda}_{12}(\hat{P}_{12}\hat{B}_{12})^{2/\alpha_{1}}+\text{ }_{2}F_{1}(-1,-\frac{2}{\alpha_{2}};1-\frac{2}{\alpha_{2}};-\theta_{2})}+\frac{1}{\hat{\lambda}_{21}(\hat{P}_{21}\hat{B}_{21})^{2/\alpha_{2}}+\text{ }_{2}F_{1}(-1,-\frac{2}{\alpha_{1}};1-\frac{2}{\alpha_{1}};-\theta_{1})},
      \end{align}
      \hrule
      \normalsize
      \end{figure*}
     The mean $M_{1,\text{total}}$ and the variance  $M_{2,\text{total}}-M_{1,\text{total}}^{2}$ can be matched to the Beta distribution of the dual-hop IoT network as follows:
     \begin{equation}\label{beta-match}
       \bar{F}_{P_{s,\text{total}}}(x)\approx 1-I_{x}\left(\frac{\beta M_{1,\text{total}}}{1-M_{1,\text{total}}},\beta\right), \quad\quad x\in [0,1],
     \end{equation}
     where $\beta\overset{\Delta}{=}\frac{(M_{1,\text{total}}-M_{2,\text{total}})(1-M_{1,\text{total}})}{M_{2,\text{total}}-M_{1,\text{total}}^{2}}$
     and $I_{x}(a,b)$ is the regularized incomplete Beta function
     \begin{equation}\label{regularized-incomplete-beta-function}
       I_{x}(a,b)\overset{\Delta}{=}\frac{\int_{0}^{x}t^{a-1}(1-t)^{b-1}d\text{t}}{B(a,b)},
     \end{equation}
     where $B(a,b)$ denotes the Beta function.

     \section{Numerical Results and Discussions}
     In this section, we first validate our results via simulations using MATLAB. We then use the developed analytical model to study the performance of the dual-hop IoT network and obtain design insights. Unless otherwise stated, we use the following parameters in our simulations and analysis. The transmission powers of tier 1 and tier 2 in the downlink direction are $P_{1}=50$ watts and $P_{2}=5$ watts, respectively. The simulated dual-hop IoT network size is $ 90\text{km}\times90 \text{km}$. We assume that the density of MBSs is $\lambda_{1}=2$ MBS/km$^{2}$ and the density of relays is $\lambda_{2}=70$ relays/km$^{2}$. The network downlink bandwidth is $10$ MHz. Transmissions by devices and relays are scheduled in a round-robin fashion in the downlink direction.

      In Fig. \ref{meta_selfbackhaul}, we validate our analytical model by depicting the exact (Gil-Pelaez) total meta distribution for a dual-hop IoT network defined in Eq. (\ref{Gil-pelaez-2hops}) and the Beta distribution of the meta distribution (\ref{beta-match}). The graphs show that the meta distributions closely match the simulation results for a wide range of $\theta$ values, confirming the accuracy of the analytical model. Fig. \ref{meta_selfbackhaul} also shows what combinations of reliability $x$ and fraction of devices can be achieved for $\theta \in \{10, 1, 0.1\}$ dB. From Fig. \ref{meta_selfbackhaul}, we note that about 23\% of the devices (when $\theta=10$), 72\% of devices (when $\theta=1$), and 98\% of devices (when $\theta=0.1$) have success probabilities equal to $0.3$.

     Fig. \ref{coverage_variance} illustrates the standard coverage success probability $M_{1,\text{total}}$ and the variance as a function of $\theta$ for the dual-hop IoT network when $B_{1}=B_{2}=1$ and $\alpha_{1}=\alpha_{2}=3$ and 4. As we can see from the graphs in Fig. \ref{coverage_variance}, the analytical results from the Beta distribution closely match the simulation results, confirming the accuracy of proposed model. Since the variance tends to zero for both $\theta \rightarrow 0$ and $\theta \rightarrow \infty$, it goes to a maximum at some finite value of $\theta$. By examining Fig. \ref{coverage_variance}, a numerical evaluation shows that for $\alpha_1= \alpha_2= 3$, the variance is maximized exactly at $\theta=0$. The coverage success probability at which the variance is maximized\footnote{\color{black}At the maximum value of variance, the coverage of devices will encounter the highest dispersion from the mean coverage probability.} is $M_{1,\text{total}} = 0.38$ for both values of $\alpha_1=\alpha_2=3$ and $\alpha_1=\alpha_2=4$.

\begin{figure}[!h]
    \begin{center}
    \scalebox{0.4}[0.4]{\includegraphics{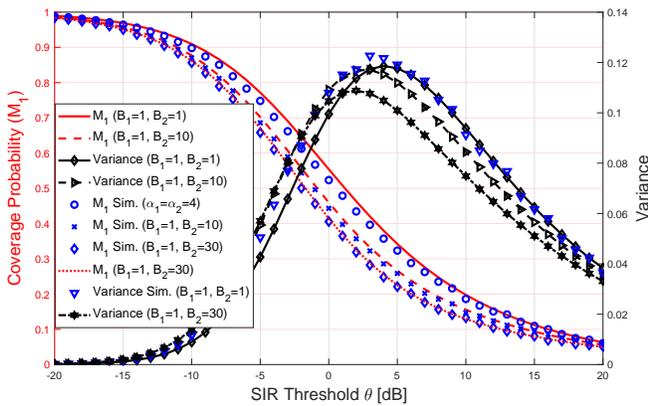}}
    %\includegraphics[width=\textwidth,keepaspectratio]{./figures/untitled.jpg}
    %\vspace{-.5cm}
    \end{center}
    \caption{Coverage probability $M_{1,\text{total}}$ and variance $M_{2,\text{total}}-M_{1,\text{total}}^{2}$ as a function of $\theta$ when $\alpha_{1}=\alpha_{2}=4$, $B_{1}=1$, and $B_{2}=1$, 10, and 30.}
     \label{coverage_variance_bias}
    \end{figure}

     In Fig. \ref{coverage_variance_bias}, we study the effect of offloading bias $B_2$ on both the entire network in terms of the first moment $M_{1,\text{total}}$, i.e., the coverage probability, and  on the variance of the conditional success probability. As we can see from Fig. \ref{coverage_variance_bias}, by offloading devices from the MBS tier to the relays tier when $B_2=10$ and $B_2=30$, the total coverage probability $M_{1,\text{total}}$ suffers a loss (this implies that $M_{1,\text{total}}$ decreases due to the increase in interference).
%\subsection{Mean local delay}\label{mean_local_delay}

     The mean local delay is the mean number of transmission attempts, i.e., re-transmissions,  for a transmitter to successfully transmit a packet to its target receiver. The mean local delay of the dual-hop IoT network is a vital performance indicator that measures the quality-of-service provided by the network, which complements the quantity-of-service, usually denoted by throughput or capacity. The mean local delay $M_{-1,\text{total}}$ of the total network, which is the $-1^{th}$ moment of the total network, can be calculated by substituting $b = -1$ into Eq. (\ref{whole-moments}). The mean local delay is defined in Eq. (\ref{local_delay_total}) shown at the top of this page.
     %Fig. \ref{coverage_variance_bias_2} depicts the total network mean local delay $M_{-1}$ as function of $\mathrm{SIR}$ threshold $\theta$. As we can see from this figure that the mean local delay goes to infinity due to the existence of a singularity at Eq. (\ref{local_delay_total}).
%\begin{figure}[!h]
%    \begin{center}
%    \scalebox{0.4}[0.4]{\includegraphics{fig/Local_delay_beta_selfbackhaul_orthog.eps}}
%    %\includegraphics[width=\textwidth,keepaspectratio]{./figures/untitled.jpg}
%    %\vspace{-.5cm}
%    \end{center}
%    \caption{Mean Local Delay $M_{-1,\text{total}}$ as a function of $\theta$ for the total communication network when $\alpha_{1}=\alpha_{2}=4$ and $B_{1}=B_{2}=1$.}
%     \label{coverage_variance_bias_2}
%    \end{figure}

\begin{figure}[!h]
    \begin{center}
    \scalebox{0.37}[0.37]{\includegraphics{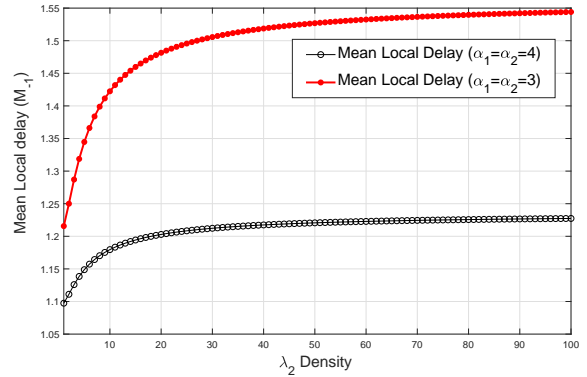}}
    %\includegraphics[width=\textwidth,keepaspectratio]{./figures/untitled.jpg}
    %\vspace{-.5cm}
    \end{center}
    \caption{The mean local delay $M_{-1,\text{total}}$ as a function of the relay density $\lambda_2$ when $\lambda_1=2\text{ MBS/KM}^{2}$, $B_{1}=1$ and $B_{2}=10$, $\alpha_{1}=\alpha_{2}=3$ and 4, and $\theta=\theta_{\mathcal{D}}=\theta_{2}=$-10 dBm.}
     \label{local_delay}
    \end{figure}

Fig. \ref{local_delay} depicts the mean local delay of the network as a function of the relay density $\lambda_2$ for the total dual-hop IoT network when $\lambda_1=2\text{ MBS/KM}^{2}$, $B_{1}=1$ and $B_{2}=10$ and $\alpha_{1}=\alpha_{2}=3$ and 4. As we can see from the the graphs, when the density of relays increases, the mean local delay of the total network increases too due to the increase in interference. However, the mean local delay of the total network in the case of $\alpha_{1}=\alpha_{2}=3$ is higher than that in the case of $\alpha_{1}=\alpha_{2}=4$. That can implies that the number of retransmissions in the case of $\alpha_{1}=\alpha_{2}=3$ is higher than that in the case of $\alpha_{1}=\alpha_{2}=4$.
\section{CONCLUSION}
The meta distribution is a fine-grained key performance metric of wireless communication systems. In this paper, we focus on downlink dual-hop IoT networks with offloading biases. We provide closed-form expressions of all moments of the conditional success probability for both relays and devices, which are then used to approximate the meta distribution using the Beta distribution. We confirm the accuracy of the approximation by simulations. We also derive the exact mean local delay in closed form for the dual-hop IoT network.

The analytical results are used to study the effect of the offloading biases on the the first moment and the variance of the conditional success probability. We study the effect of increasing the relay density on the performance of a typical device and
other devices in terms of the mean local delay and the meta distribution. The mean local delay of the dual-hop IoT network increases as the relay density increases.

\appendices
\section{Proof of Lemma \ref{1}}\label{prof_1}
While taking the offloading biases effect in consideration, the $b^{th}$ moment of the CSP $P_{s,k}(\theta_{\mathcal{D}})$ of the typical device when it is served by the $k^{th}$ tier is given as follows:
\vspace{-0.4cm}
 \begingroup
    \allowdisplaybreaks
\footnotesize
\begin{flushleft}
\begin{align}\label{MBS-or-SBS-b-moment-2}
            M_{b,\text{k}}&=\mathbb{E}_{d_{k,\mathcal{D}}}\bigg[\mathbb{P}(n=k|d_{k,\mathcal{D}})P_{s,k}(\theta_{\mathcal{D}})^{b}\bigg],\notag\\
                 &\stackrel{(a)}{=}\mathbb{E}_{d_{k,\mathcal{D}}}\bigg[\prod_{j\neq k}e^{-\pi\lambda_{j}(\hat{P}_{jk}\hat{B}_{jk})^{2/\alpha_{j}}r^{2}}\times\notag\\
                 &\quad\quad\quad\prod_{\mathbf{y}_{k,i}\in\mathbf{\Phi}_k\backslash \{\mathbf{y}_{k,0}\}}\frac{1}{\left(1+\theta_{\mathcal{D}}\left(\frac{d_{k,\mathcal{D}}}{\|\mathbf{y}_{k,i}\|}\right)^{\alpha_{k}}\right)^{b}}\bigg],\notag\\
                 &\stackrel{(b)}{=}\mathbb{E}_{d_{k,\mathcal{D}}}\bigg[\prod_{j\neq k}e^{-\pi\lambda_{j}(\hat{P}_{jk}\hat{B}_{jk})^{2/\alpha_{j}}r^{2}}\times\notag\\
                 &\quad\quad\quad\exp
                 \left(\int_{d_{k,\mathcal{D}}}^{\infty}-2\lambda_{k}\pi\bigg[1-\frac{1}{\left(1+\theta_{\mathcal{D}}\left(\frac{d_{k,\mathcal{D}}}{y}\right)^{\alpha_{k}}\right)^{b}}\bigg]y\text{d}y\right)\bigg],\notag\\
                 &\stackrel{(c)}{=}\int_{0}^{\infty}2\lambda_{k}\pi r e^{-\lambda_{k}\pi r^{2}}e^{-\sum\limits_{j\neq k}\lambda_{j}(\hat{P}_{jk}\hat{B}_{jk})^{2/\alpha_{j}}\pi r^{2}}\times\notag\\
                 &\quad\quad\quad\exp\left(\int_{r}^{\infty}-2\lambda_{k}\pi\bigg[1-\frac{1}{\left(1+\theta_{\mathcal{D}}\left(\frac{r}{y}\right)^{\alpha_{k}}\right)^{b}}\bigg]y\text{d}y\right)\text{d}r,\notag\\
                 &\stackrel{(d)}{=}\int_{0}^{\infty}e^{-q} e^{-q\sum\limits_{j\neq k}\hat{\lambda}_{jk}(\hat{P}_{jk}\hat{B}_{jk})^{2/\alpha_{j}}}\times\notag\\
                 &\quad\quad\quad\exp\left(-2q\int_{0}^{1}\bigg[1-\frac{1}{\left(1+\theta_{\mathcal{D}}v^{\alpha_{k}}\right)^{b}}\bigg]v^{-3}\text{d}v\right)\text{d}q,\notag\\
                 &\stackrel{(e)}{=}\int_{0}^{\infty}e^{-q} e^{-q\sum\limits_{j\neq k}\hat{\lambda}_{jk}(\hat{P}_{jk}\hat{B}_{jk})^{2/\alpha_{j}}}\times\notag\\
                 &\quad\quad\quad\exp\left(-q\int_{1}^{\infty}\bigg[1-\frac{1}{\left(1+\theta_{\mathcal{D}}u^{-\alpha_{k}/2}\right)^{b}}\bigg]\text{d}u\right)\text{d}q,\notag\\
                 &\stackrel{(f)}{=}\int_{0}^{\infty}e^{-q} e^{-q\sum\limits_{j\neq k}\hat{\lambda}_{jk}(\hat{P}_{jk}\hat{B}_{jk})^{2/\alpha_{j}}}\times\notag\\
                 &\quad\quad\quad\exp\left(-q\bigg[\text{ }_{2}F_{1}(b,-\frac{2}{\alpha_{k}};1-\frac{2}{\alpha_{k}};-\theta_{\mathcal{D}})-1\bigg]\right)\text{d}q,\notag\\
                 &=\int_{0}^{\infty} e^{-q\bigg[\sum\limits_{j\neq k}\hat{\lambda}_{jk}(\hat{P}_{jk}\hat{B}_{jk})^{2/\alpha_{j}}+\text{ }_{2}F_{1}(b,-\frac{2}{\alpha_{k}};1-\frac{2}{\alpha_{k}};-\theta_{\mathcal{D}})\bigg]}\text{d}q,\notag\\
                 &=\frac{1}{\sum\limits_{j\neq k}\hat{\lambda}_{jk}(\hat{P}_{jk}\hat{B}_{jk})^{2/\alpha_{j}}+\text{ }_{2}F_{1}(b,-\frac{2}{\alpha_{k}};1-\frac{2}{\alpha_{k}};-\theta_{\mathcal{D}})}.
          \end{align}
\end{flushleft}
\normalsize
\endgroup

where (a) follows from considering the conditional association probability for the typical device connecting to the $k^{th}$ tier given in Eq. \eqref{condition-access}. In step (b) we apply PGFL of the PPP \cite[Chapter 4]{haenggi2012book}. Step (c) follows from averaging over $d_{k,\mathcal{D}}$, step (d) is by using variable substitution $q=\pi\lambda_{k}r^{2}$ and $v=r/y$. In step (e) we perform variable substitution $v=u(\hat{P}_{jk}\hat{B}_{jk})^{-1/\alpha_{j}}$ and step (f) follows from the fact that $\text{ }_{2}F_{1}(b,-\frac{2}{\alpha};1-\frac{2}{\alpha};-\theta)\equiv 1+\int_{1}^{\infty}(1-\frac{1}{(1+\theta h^{-\alpha/2})^{b}})\text{d}h$.

%\label{conc_future}
%We present a .....
%\clearpage

%\appendices

%\section{Proof of Lemma \ref{pdf_distance}}
%\label{prof_3}
%
%\section{Proof of Lemma \ref{average_trans_rate_three_types}}
%%%============================================================
%\label{Average transmission rate_prof}


\ifCLASSOPTIONcaptionsoff
 % \newpage
\fi

%\newpage
\bibliographystyle{IEEEtran}
\bibliography{ICC-2019}
%\begin{thebibliography}{1}
%\end{thebibliography}

\end{document}